\documentclass[twocolumn,amsmath,floats,showpacs,nofootinbib]{revtex4}
\usepackage[usenames]{color}
\usepackage{graphicx}
\usepackage{epstopdf}
\usepackage{textcomp}
\usepackage{hyperref}
\usepackage{multirow}
\usepackage{tikz}
\usepackage{rotating}
\hypersetup{colorlinks=true}

\begin{document}

\title{Electron-quasiparticle interaction in $\rm DyNi_2B_2C$ measured by point-contact spectroscopy}

\author{I.K. Yanson$^a$, N.L. Bobrov$^a$, C.V. Tomy$^{b,c}$, D.McK. Paul$^b$}
\affiliation{$^a$B.Verkin Institute for Low Temperature Physics and Engineering, 47, Lenin Ave., 310164 Kharkov, Ukraine\\
$^b$Department of Physics, University of Warwick, Coventry CV4 7AL, UK\\
$^c$Department of Physics, Indian Institute of Technology, Powai, Mumbai 400 076, India\\
Email address: bobrov@ilt.kharkov.ua}
\published {\href{http://dx.doi.org/10.1016/S0921-4534(00)00226-4}{Physica C} 334 (2000) 152-162}
\date{\today}

\begin{abstract}The electron-quasiparticle interaction (EQI) spectral function has been measured for $\rm DyNi_2B_2C$ in the normal state at low temperatures by means of point-contact spectroscopy (PCS). A low-frequency peak is found around $eV\sim 5\ meV$. It becomes measurable at $T_m^*\simeq 15\ K$ and grows in intensity with constant width as the temperature is lowered. We argue that this peak arises from the strong interaction of conduction electron with coupled crystal-electric-field-phonon excitations whose branches cross at low energy. The comparison with PC spectra for $\rm HoNi_2B_2C$ suggests that a similar peak also exists for this compound. The magnitude of the point-contact EQI parameter $\lambda_{PC}$ for $\rm DyNi_2B_2C$ is estimated.  $\copyright$2000 Published by Elsevier Science B.V. All rights reserved.\\

\end{abstract}

\maketitle
\section{INTRODUCTION}
Among the $\rm RNi_2B_2C$-family of magnetic superconductors (R is Tm, Ho, Er, or Dy), $\rm DyNi_2B_2C$ is the only material whose superconducting transition temperature $T_c = 6.2\ K$ is well below its Neel transition temperature $T_N = 10.5\ K$. In the absence of an applied magnetic field, it behaves as a conventional superconductor \cite{1} due to the commensurate antiferromagnetic (AFM) order, although perhaps with a degree of anisotropy in the superconducting energy gap. Besides the averaged superconducting energy gap which satisfies the BCS relation, de Haas-van Alphen measurements have found a small superconducting gap on part of the Fermi surface of $\rm YNi_2B_2C$ which is a nonmagnetic analogue of $\rm DyNi_2B_2C$ \cite{2}. As the other $\rm RNi_2B_2C$, $\rm DyNi_2B_2C$ has the body- centered tetragonal crystal structure \cite{3} and its Fermi surface possesses strong nesting for wave vectors in the $a$-direction \cite{4} which for the Y- and  $\rm LuNi_2B_2C$ has the value (0.45,0,0). Nevertheless, unlike the Er- and Ho-compounds\footnote{In Ho-compound, an $a$ modulation only exists over a very small temperature range.}, in $\rm DyNi_2B_2C$ no magnetic ordering along $a$-direction is observed \cite{3}.

The electron-phonon interaction in this group of materials is quite strong and anisotropic which is demonstrated by a softening of the low frequency phonon modes with decreasing the temperature, leading to the appearance of a new narrow phonon line at about $4\ meV$ in nonmagnetic Y- and Lu-compounds at $T < T_c$ \cite{5,6}. In spite of previous suggestions about the vital importance of high frequency ($100\ meV$) boron mode for mediating Cooper pairing, it was shown \cite{7,8,9,10} that the low frequency phonon modes (4-15~$meV$) are more important for the
superconductivity mechanism. Especially interesting is the lowest-lying phonon mode at several $meV$ whose intensity is strongly temperature- and magnetic field-dependent. In Y- and Lu-compounds, this mode appears to correlate with the superconducting transition \cite{5,11,12}. Although in Refs. \cite{7,8,9} the remanent of this mode in $\rm YNi_2B_2C$ was observed in PC spectra without any traces of superconducting energy gap\footnote{That is why its appearance was ascribed to the normal state \cite{7,8,9}.}, it could also be ascribed to superconductivity which was not completely destroyed at fields not far from $H_{c2}$.

At low energies the crystal electric field excitations (CEF) are observed in magnetic $\rm RNi_2B_2C$. In Ho-compound they are studied by fitting the temperature-dependent magnetic susceptibility of singlecrystal \cite{13}, and in Ho-, Er-, and Tm-compounds by inelastic neutron scattering \cite{14}. Their energies amount several $meV$, and the conduction electrons interact with them strongly.

In the present work, we find that a similar strong and narrow peak in the electron-quasiparticle interaction (EQI) at about few $meV$ arises for the Dy- compound ($\simeq 5\ meV$), starting at a temperature as high as 15~$K$, which is far above the superconducting transition. This peak presumably has a common origin with CEF excitations and magnetic ordering in this compound. In $\rm HoNi_2B_2C$, at the temperature also higher than the transition to the magnetic and superconducting states, a similar peak occurs ($\simeq 12\ K$) \cite{8}, whose intensity is strongly dependent on magnetic field. Hence, we see that not only the interaction with superconducting order leads to the appearance of strong and narrow phonon mode in the EQI function, but that a same effect may occur from an interaction with the magnetic correlations. The cause for the two may be the strong interaction of conduction electrons with CEF and phonon branches of quasiparticle excitations.
\section{Method}
Point-contact spectroscopy (PCS) involves studies of nonlinearities in the current-voltage characteristics of metallic constrictions with characteristic size $d$ smaller that the inelastic electron mean free path $l_{in}$. At low temperatures, the voltage dependence of the differential resistance $R = dV/dI(V)$ of a ballistic point contact reflects the energy-dependence of the scattering cross-section of the conduction electrons off any energy dependent scatterers. Consequently, the point-contact spectra [$d^2V/dI^2(V)$ dependences] are proportional to the EQI spectral function. In the case of point contacts between dissimilar metals with very different Fermi velocities $v_F$, only the spectrum of the material with smaller $v_F$ is seen \cite{15}:
\begin{equation}
\label{eq__1}
{{\left. \frac{d\ln R}{dV}(V)=\frac{4}{3}\frac{ed}{\hbar {{v}_{F}}}{{g}_{PC}}(\omega ) \right|}_{\hbar \omega =eV}}(T\simeq0).
\end{equation}

The function $g_{PC} (\omega) = \alpha_{PC}^2 F(\omega)$ is similar to the Eliashberg function. $\alpha_{PC}^2 F(\omega)$ is the averaged EQI matrix element with kinematic restrictions imposed by the contact geometry and $F(\omega)$ is the quasiparticle (i.e. phonon) density of states. If the movement of charge carriers through the contact is diffusive ($l_e\ll d$), then the contact size $d$ in formula (\ref{eq__1}) should be replaced by the elastic mean free path $l_e$ provided the spectroscopic condition is fulfilled:
\begin{equation}
\label{eq__2}
d\le \sqrt{{{l}_{e}}{{l}_{in}}}\quad .
\end{equation}
The contact diameter $d$ is determined by the normal-state resistance at zero bias $R_0$ via the Sharvin expression in the ballistic regime. Our contacts are made between single crystal $\rm DyNi_2B_2C$, nominally within the $ab$ plane, and Ag. In the case homocontact made of Ag $d\simeq 27/ \sqrt{R_0}\ nm$, while for $\rm RNi_2B_2C$ $d\simeq 24/ \sqrt{R_0}\ nm$. Here we take for
$\rm DyNi_2B_2C$ as for the Lu-compound $N(0) = 4.8$ [states/$eV$ unit cell] and $v_F = 3.6\times 10^7\ cm/s$ for $\rho l={\frac{3}{2}}/{{{e}^{2}}N(0)}\;$ \cite{16}. For contacts with dissimilar electrodes, $\rm DyNi_2B_2C-Ag$, we shall use $d\simeq 25.5/ \sqrt{R_0}\ nm$ for further estimates since indirect evidence points to ballistic electron flow.

In the spectroscopic regime no heating of the contact occurs. However, if the contact size is large compared with the electron energy-relaxation length, then there is local heating with the maximum temperature $T_0$ given by the Kohlraush relation:
\begin{equation}
\label{eq__3}
V^2=4L(T_0^2-T^2)
\end{equation}

for an applied voltage $V$ across the contact at the bath temperature $T$. For a typical Lorentz number $L$ in the thermal regime the Einstein oscillator $\hbar\omega_0$ looks like a
smeared step with a flat maximum at $eV = 1.09\hbar\omega_0$ in the PC EQI spectrum \cite{8}. The thermal feature in the PC spectra can be quite sharp if a phase transition occurs at a particular temperature $T_m$, which leads to the jump-like increase in the temperature-dependent contribution to the resistivity. Such a situation holds in the $\rm HoNi_2B_2C$ contacts due to the magnetic transition \cite{8}, and, as we shall see below, in $\rm DyNi_2B_2C$ as well.

In the experiment we recorded the second derivative signal $V_2$ vs. the bias voltage $V$. Knowing the modulation signal $V_1$ and contact resistance $R_0$ one can write the expressions \cite{15}:
\[\frac{{{d}^{2}}V}{d{{I}^{2}}}(V)\simeq 2\sqrt{2}{{\left( \frac{{{R}_{0}}}{{{V}_{1}}} \right)}^{2}}{{V}_{2}}(V)\ \ \ \text{and}\]
\begin{equation}
\label{eq__4}
\frac{d\ln R}{dV}=2\sqrt{2}\frac{{{V}_{2}}(V)}{{{\left( {{V}_{1}} \right)}^{2}}}
\end{equation}
where $V_1,\ V_2$ in $[V]$ are the first and second harmonics of the modulation signal (rms), and $R$ is differential resistance in $[\Omega]$. $R_0$ stands for resistance at zero bias.

Unfortunately, the application of a magnetic field mechanically destroys the contact, probably because of large magnetostriction. This effect hinders the study of the influence of magnetic field on the PC spectra.

\begin{figure}[]
\includegraphics[width=8.5cm,angle=0]{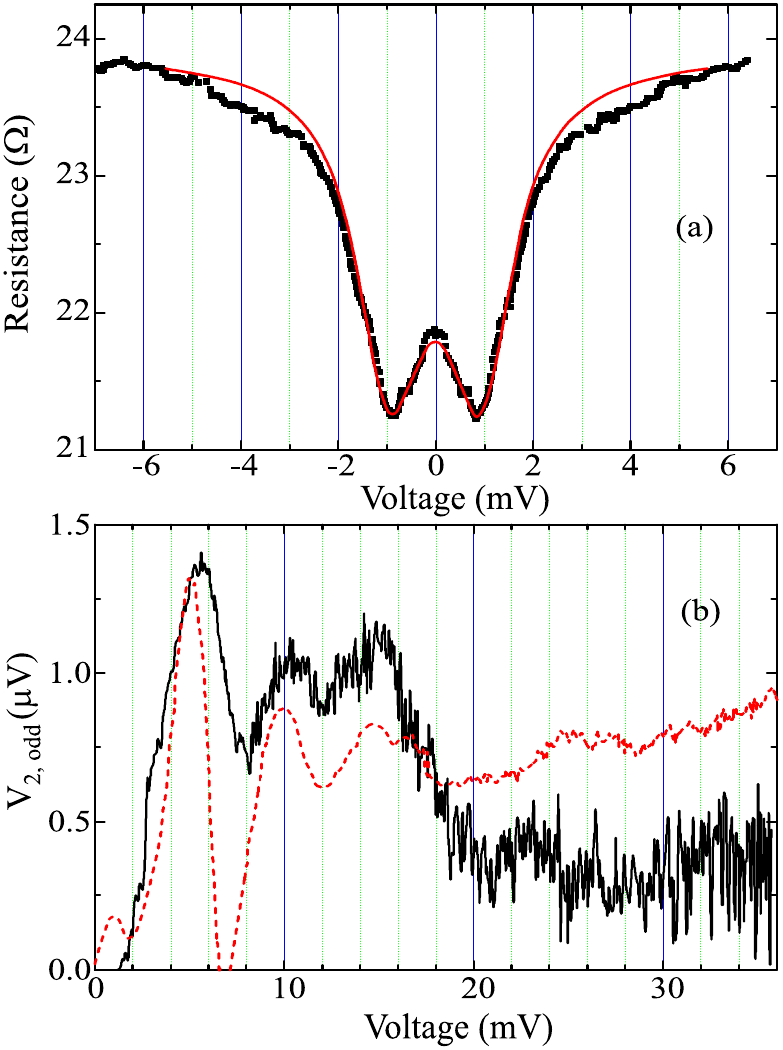}
\caption[]{(a) Differential resistance of the $\rm DyNi_2B_2C-Ag$ contact (dots) $(R_0 = 21.7\ \Omega)$ with the BTK-fit (solid line) with parameters $\Delta= 1.14\ meV$, $Z=0.40$, $\Gamma= 0.24\ meV$,  $T=1.75\ K$. (b) Odd part of the point-contact spectra for two $\rm DyNi_2B_2C$ contacts in the normal state with $R_0 = 21.7\ \Omega$, $V_{1,0} = 1.6\ mV$ (solid line, the same contact as in (a)) and $2.5\ \Omega$, $V_{1,0} = 0.88\ mV$ (dashed line), respectively. $T = 1.72\ K$, $H = 0.91\ T$. Everywhere in the paper the field is applied nominally within $ab$ plane.}
\label{Fig1}
\end{figure}

For a detailed investigation we choose contacts which show clear evidence of a superconducting phase on the surface of $\rm DyNi_2B_2C$ single crystal \cite{1}. This is indicated by a double minimum in the $R(V)$ curve due to Andreev reflection at $T\ll T_c$ when fitting to the Blonder-Tinkham-Klapwijk (BTK) procedure \cite{17}. Such a fit yields the superconducting energy gap $\rm \Delta_0( DyNi_2B_2C)\simeq 1\ meV$ and critical temperature about 6.2~$K$ (see Fig. \ref{Fig1}(a)). Since the superconducting phase is extremely sensitive to composition and crystal structure, it gives us evidence that the material under the contact is intact. The small (compared to the energy gap) value of $\Gamma$ stands for the depairing interaction inside the contact
and does not influence the value of the superconducting energy gap.

The PCS spectrum of these junctions in the superconducting state should not contain sharp peaks in the first derivative $R(V)$ which look like an N-type singularity in the PC spectra $V_2(V)$ and whose position on a voltage scale strongly depends on the junction resistance. This singularity points to the nonequilibrium destruction of superconductivity due to heating or an excess concentration of normal quasiparticles. In the normal state, at magnetic fields greater than $H_{c2}(0)\sim 0.7\ T$ \cite{18}, the positions of the spectral peaks also should not depend upon the contact resistance except the possible change in shape. The shape of the peak can alter between normal Gaussian peaks, as in any PC spectra, and N-type feature which is equivalent to the maximum in $R(V)$. The positions of this N-type feature depends on the destruction of AFM order and, consequently, to the step-like change in resistivity at this particular bias. An example of such spectra are plotted in Fig. \ref{Fig1}(b) where the contact with resistance of $22\ \Omega$ shows the conventional peak shape while that of $R = 2.5\ \Omega$ transforms to the N-shape form (which is equivalent to the maximum in $dV/dI$) located at approximately the same bias voltage of about $5\ meV$. The latter is due to the "quasi-thermal" regime discussed in Ref. \cite{18} for Ho-compound. Namely, due to a strong phonon-magnon interaction the effective boson subsystem temperature increases up to the magnetic transition which leads to the step-like increase in resistivity of the metal under the contact and consequently to a N-type singularity in the PC spectra.

\begin{figure}[]
\includegraphics[width=8.5cm,angle=0]{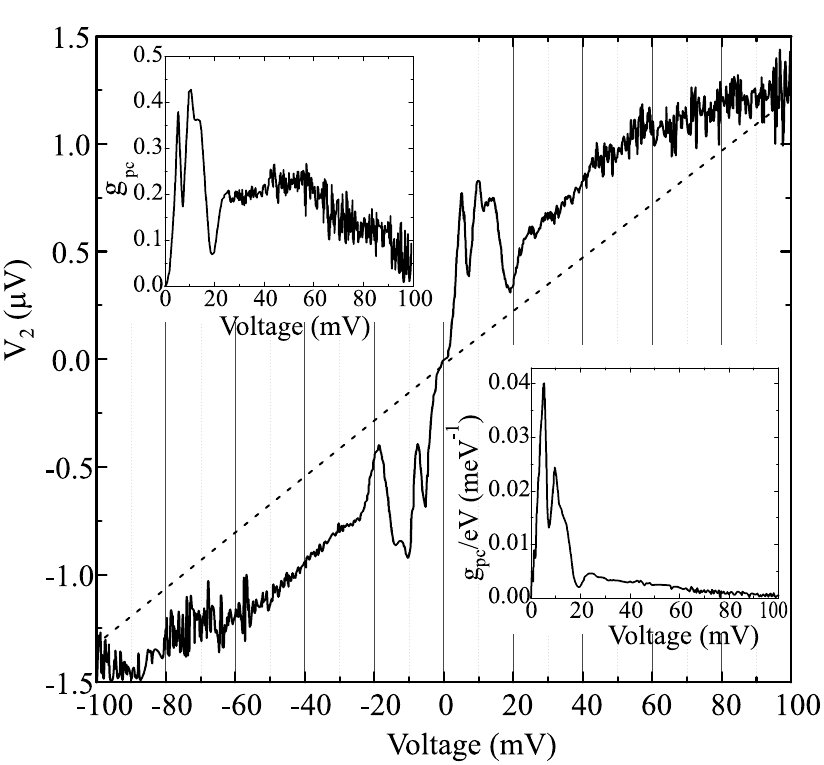}
\caption[]{Point-contact spectrum ($V_2$-second harmonic of modulation signal) of $\rm DyNi_2B_2C-Ag$ contact with $R_0=27\ \Omega$, $V_{1,0}=0.587\ mV$, $T=1.83\ K$, $H=0.65\ T$. The dashed straight line indicates the supposed background. The upper inset plots the spectrum minus background in units of $g_{PC}$ (see Eq.(\ref{eq__1})). The lower inset shows the function $g_{PC}/eV$ which enters into the integral for $\lambda_{PC}$ (Eq.(\ref{eq__6})). This contact shows the maximum value of $\lambda_{PC}=0.85$ calculated for a clean orifice model (see Eq.(\ref{eq__1})).}
\label{Fig2}
\end{figure}

Normally, the PCS spectra are approximately odd with respect to applied voltage. To show this, we present the spectra swept from $-V_{max}$ to $+V_{max}$ (see Figs. \ref{Fig2} and \ref{Fig4}). If there is some (small) asymmetry, we plot the odd part of second harmonic:
\begin{equation}
\label{eq__5}
{{V}_{2,\text{odd}}}=\frac{1}{2}\left[ {{V}_{2}}(V)-{{V}_{2}}(-V) \right]\quad .
\end{equation}

These PC spectra contain a smoothly rising background. For a high-resistance junction the background can be quite low (see Fig.\ref{Fig1}(b) for the contact with $R= 22\ \Omega$), while for low-ohmic junctions it becomes a large percentage of the signal, especially at high biases. For simplicity, we approximate the background by a straight line (shown, for example, in Fig.\ref{Fig2} by a dashed line) in analogy with previous work on similar contacts for $\rm YNi_2B_2C$ and $\rm HoNi_2B_2C$ (see Ref.\cite{8}).

\section{Results}
\subsection{Determination of $\lambda_{PC}$}
\begin{figure}[]
\includegraphics[width=8.5cm,angle=0]{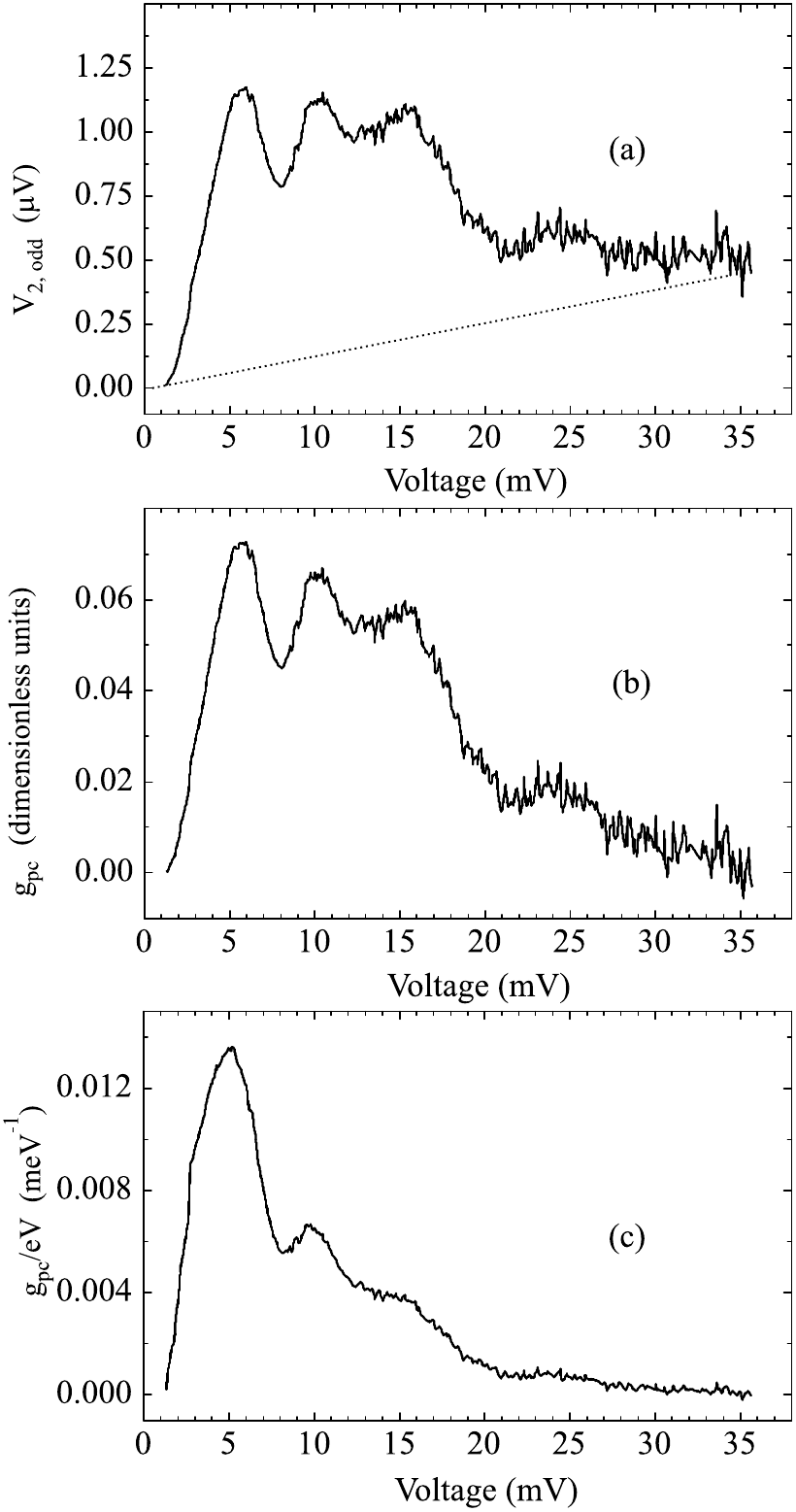}
\caption[]{Odd part of the PC spectrum (a), $g_{PC}(eV)$ (b) and $g_{PC}(eV)/eV$ (c) functions for typical high-ohmic contact, $\rm DyNi_2B_2C-Ag$. $R_0=21.5\ \Omega$ $(d=5.4\ nm$), $V_{1,0}=1.19\ mV$, $T = 1.72\ K$ and $H = 0.91\ T$. $\lambda_{PC} = 0.23$ for the model of clean orifice (Eq.\ref{eq__1}). All the peaks have a Gaussian form without "undershooting". The background is supposed to be the dotted straight line shown on the upper panel.}
\label{Fig3}
\end{figure}

One of the best quality PC spectra for $\rm DyNi_2B_2C-Ag$ contacts is shown in Fig.\ref{Fig2}. Large intensity low-lying peaks in $V_2(V)$ are clearly seen at biases $5-15\ meV$. In addition, there are flat maxima at 25, 45, 55~$meV$ which can be more clearly seen by subtracting the linear background. This is done in the upper inset of Fig.\ref{Fig2}. $g_{PC}(eV)$ can be estimated from $V_2(V)$ using Eqs.(\ref{eq__1}) and (\ref{eq__4}). According to Ref.\cite{10}, the phonon spectrum for $\rm RNi_2B_2C$ should stretch to 160~$meV$ with boron $A_{1g}$ modes located at  $\simeq 100\ meV$. In theoretical study, Ref.\cite{19}, this mode is considered as a primary source for the Cooper pair interaction. The PC spectra of EQI of $\rm DyNi_2B_2C$ (see Fig.\ref{Fig2}) as well as those for $\rm YNi_2B_2C$, $\rm HoNi_2B_2C$, and $\rm LaNi_2B_2C$ \cite{7,8,9} saturate at biases of the order of $100\ meV$ which point to a considerable electron-phonon interaction at this energy. Nevertheless, no differences were observed between superconducting ($\rm YNi_2B_2C$ and $\rm HoNi_2B_2C$) and nonsuperconducting ($\rm LaNi_2B_2C$) compounds at biases  $\simeq 100\ meV$, while large differences were found at lower energies. The authors of Ref.\cite{10} come to a similar conclusion by comparing neutron spectra of phonons for superconducting and nonsuperconducting pseudoquaternary compounds of the form $\rm Y(Ni_{1-x}Co_x)_2B_2C$. Indeed, the electron-phonon-interaction parameter is equal:
\begin{equation}
\label{eq__6}
{{\lambda }_{PC}}=2\int\limits_{0}^{\infty }{({{{g}_{PC}}}/{eV)d(eV)}\;}
\end{equation}
and in order to have a comparable influence to the strong peaks at about 5-10~$meV$, the high-frequency boron peak should be at least an order of magnitude
more intense. This is not the case as seen from the lower inset of Fig.\ref{Fig2} which plots the $g_{PC}/eV$-function. An estimate of $\lambda_{PC}$ gives for the junction of Fig.\ref{Fig2} the value of 0.85\footnote{In the review \cite{9}, where the same $\rm DyNi_2B_2C-Ag$ contact was considered, an extra factor of 2 mistakenly was not taken into account in the estimation of $\lambda_{PC}$.}  which is the highest value measured in our experiments. One should be cautious with estimates of $\lambda_{PC}$. It can be greater than the true electron-phonon-interaction parameter if some "quasi-thermal" processes are involved, due to the destruction of magnetic and/or superconducting order. On the other hand, $\lambda_{PC}$, determined using formula (\ref{eq__1}) for the ballistic regime \cite{15}, can also be substantially smaller than the true value, because of the above-mentioned diffusive motion of charge carriers through the contact. In the present example (Fig.\ref{Fig2}), the possible "undershooting" of the spectrum at $eV=20\ meV$ may hint that a "quasi-thermal" model is
appropriate. More typical is the spectrum shown in Fig.\ref{Fig3}, where no "undershooting" occurs explicitly. For an approximate estimation of $\lambda_{PC}$, the bias range up to the highest phonon energy (160~$meV$) does not matter, since we have seen from the lower inset of Fig.\ref{Fig2} that the higher-bias phonon structure does not influence the magnitude of the integral (Eq.\ref{eq__6}). Here, the value of $\lambda_{PC}$ is equal to 0.23 which is a more typical magnitude for a number of the observed PC spectra of $\rm DyNi_2B_2C$. To compare the typical length scales, for contacts with $R=27(21.5)\ \Omega$, the Sharvin formula gives the diameter $d=4.9(5.5)\ nm$ while the electron mean free path at $T\geq T_c$ for $\rm DyNi_2B_2C$, according to Ref.\cite{16}, equals $19.5\ nm$ within the $ab$ plane for the best single crystal. Since we are not quite sure that the mean free path under the contact equals the value above, we can only assume that the regime of current flow corresponds to the ballistic flow.
\subsection{Temperature dependence of the lowest-lying peak}

\begin{figure}[]
\includegraphics[width=8.5cm,angle=0]{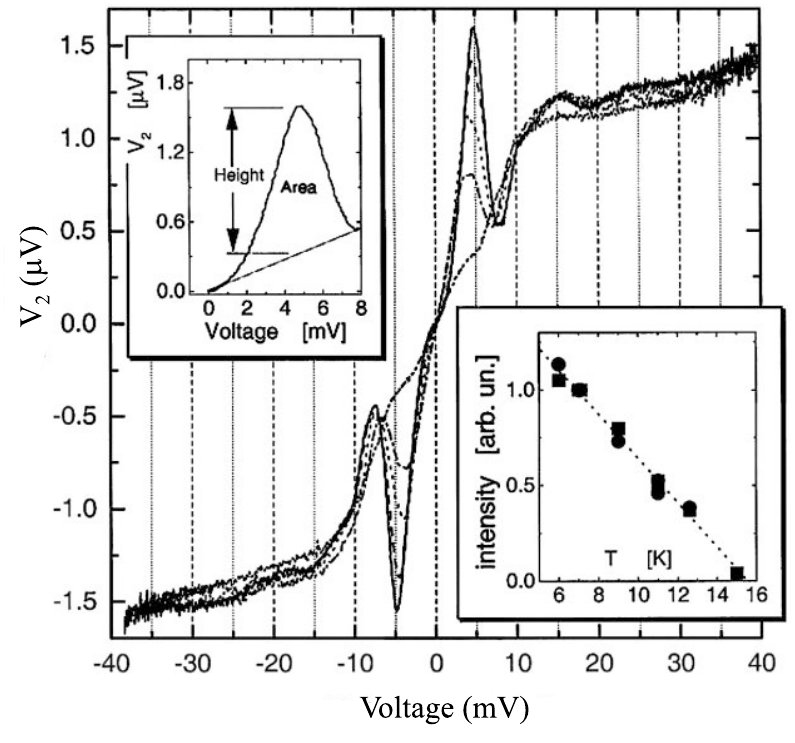}
\caption[]{The central result of the paper: The temperature dependence of the PC spectra in the normal state for the contact $\rm DyNi_2B_2C-Ag$ with $R_0=4.8\ \Omega$, $V_{1,0} = 0.76\ mV$. $T$=6, 7, 9, 11 and 15 $K$ for different curves, respectively. For the intensity of the lowest-lying peak, either the area or the height separately, defined as is shown in the upper inset for $T=6\ K$, are used. The lower inset displays the temperature dependence of these parameters (squares-area, dots-heights) normalized at $T=7\ K$. Three more pairs of dots are added corresponding to the normal state spectra of the contact whose characteristic is shown in Fig.\ref{Fig5}. Note the onset temperature of $T_m^*=15-16\ K$ which is substantially higher than $T_N=10.5\ K$. All the curves are taken at zero magnetic field.}
\label{Fig4}
\end{figure}

The central result of this work is the pronounced temperature dependence of the lowest-lying peak in the EQI function of $\rm DyNi_2B_2C$ (Fig.\ref{Fig4}). In Ref.\cite{8} it was already noticed that the lowest-lying peak in $\rm HoNi_2B_2C$ appeared at a temperature of $\sim 12\ K$, i.e. higher than the superconducting transition temperature (8.5~$K$) and incommensurate spin-wave-transition temperature along the $c$-axis (also about $8.5\ K$).\footnote{This coincidence of transition temperature for both types of long-range orders hinders their separation through the Andreev-reflection measurements in Ref.\cite{20}.}

Unlike the phonon structure at biases 10-25~$meV$, which is almost independent on temperature in Fig.\ref{Fig4} (besides some loss of clarity due to thermal smearing), the peak at 5~$meV$ strongly depends on temperature. If we define its height and area as shown in the upper inset of Fig.\ref{Fig4}, then its intensity $I$, determined by both these two quantities, follows the line shown in the lower inset of Fig.\ref{Fig4}. This shows that its width is almost independent of temperature. The peak position shifts to somewhat lower values with increasing temperature similarly to the lowest-lying peak in Ho-compound \cite{8} (see, also, Fig.\ref{Fig8}). We note that similar width and position were
observed for the lowest-lying peak in the nonmagnetic (but superconducting) $\rm LuNi_2B_2C$ compound as well, according to the neutron studies (see Fig.5 of Ref.\cite{11}). The position of neither of these anomalous peaks extrapolates to zero at the onset temperatures (15, 12 and 16 $K$, respectively).
\begin{figure}[]
\includegraphics[width=8.5cm,angle=0]{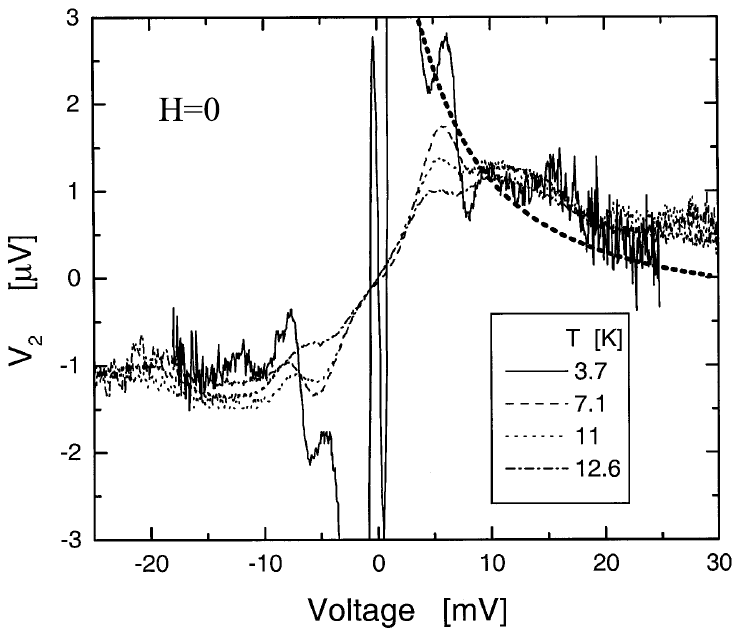}
\caption[]{The temperature dependence of PC spectra for another $\rm DyNi_2B_2C-Ag$ contact including the spectrum in the superconducting state ($T=3.7\ K$, shown with a straight line). Note the strong nonlinearities near zero-bias caused by the superconducting energy gap. For higher biases, these nonlinearities tend to the background which is schematically drawn with a thick dotted curve only for positive biases. It is seen that the lowest-lying peak has an "undershooting" feature with respect to this background. For higher energies, the normal and superconducting spectra approximately coincide disregarding the larger noise in the superconducting state. Parameters of the contact are $R_0=15.8\ \Omega$ except for $T = 3.7\ K$ where a jump to $R_0=22\ \Omega$ occurs, $V_{1,0}=1.5\ mV$.}
\label{Fig5}
\end{figure}

The lowest-lying peak in $\rm DyNi_2B_2C$ starts at $T=15\ K$ and increases almost linearly in intensity with temperature down to $T\sim T_c$, the superconducting transition temperature. Note that neither the transition to the AFM state at $T_N=10.5\ K$, nor to the superconducting state at $T_c=6.2\ K$ dramatically influences the $I(T)$-dependence. Fig.\ref{Fig5} shows the spectra at different temperatures for another junction, where the spectrum in the superconducting state (at $T= 3.7\ K$) is shown together with those in the normal state. If one subtracts the steeply rising background due to the energy gap nonlinearity at lower biases (which is thought to look like the thick dotted curve for the positive biases in Fig.\ref{Fig5}), then the $5\ meV$ peak intensity remains approximately the same as for the normal state at $T\geq T_c$. Here, the quasithermal "undershooting" is seen. In this case, the "undershootin"  is caused by the suppression of superconductivity as was studied in detail previously for phonon structure in Ta junctions \cite{21}. Note that the phonon structure at 10-25~$meV$ remains almost the same both in the normal and superconducting states, as it should be for weak coupling superconductivity \cite{22,23}.
\section{Discussion}
\begin{figure}[]
\includegraphics[width=8.5cm,angle=0]{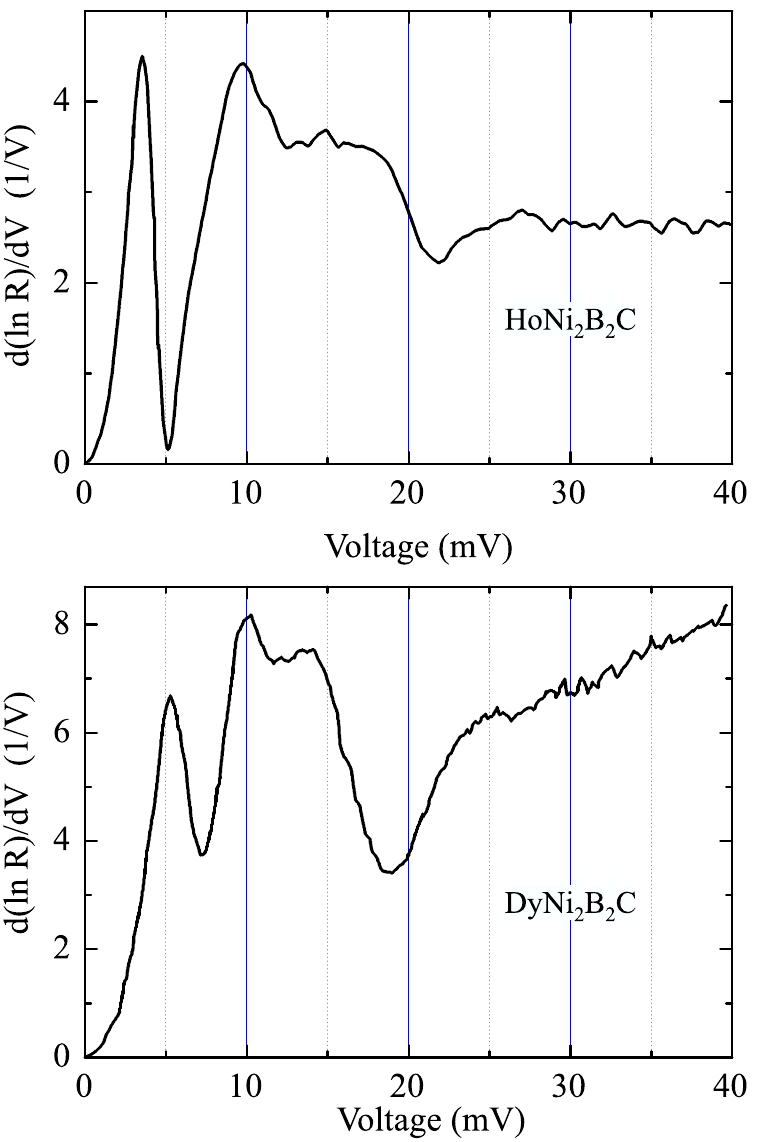}
\caption[]{Comparison between the PC spectra of $\rm HoNi_2B_2C-Ag$ and $\rm DyNi_2B_2C-Ag$ contacts in the normal state. A general similarity in the shape is seen. The parameters for Ho-system: $R_0 = 2.3\ \Omega$, $H = 0.5\ T$, $T = 4.2\ K$; for Dy-system: $R_0 = 27\ \Omega$, $H = 0.65\ T$, $T = 1.8\ K$.}
\label{Fig6}
\end{figure}

Let us compare the PC spectra for Ho- and Dy- compounds in the normal state obtained at low temperatures ($T\ll T_c$) by applying the requisite magnetic field to destroy the superconductivity (Fig.\ref{Fig6}).

\begin{figure}[]
\includegraphics[width=8.5cm,angle=0]{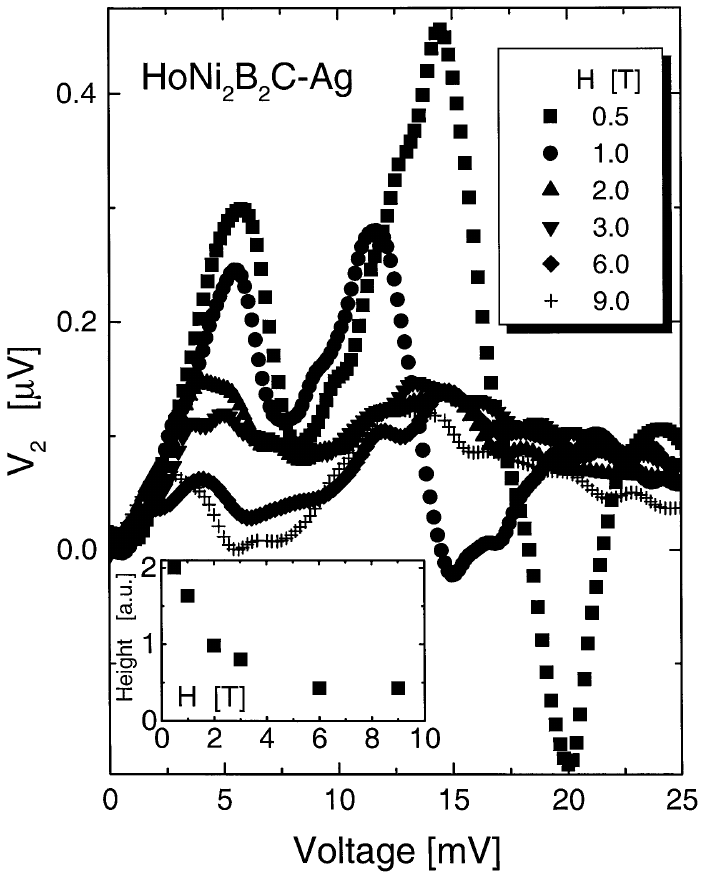}
\caption[]{PC spectra for different magnetic fields H for $\rm HoNi_2B_2C-Ag$ contact replotted from Fig.7 of Ref.\cite{8} linear background being subtracted. $R_0 = 0.47\ \Omega$, $V_{1,0} = 0.8\ mV$, $T = 4.2\ K$. The lowest-lying peak has no undershooting features. Its height vs. magnetic field is shown in the inset.}
\label{Fig7}
\end{figure}

These compounds possess the same low-temperature commensurate AFM state (AF-I-related, with ordered moments aligned along [110] direction). Despite the fact they reach its ground state through a different succession of magnetic structures, their PC spectra are very similar. Both of them consist of a lowest-lying peak (in Ho-compound at 3.5~$meV$ and in Dy-compound at 5.5~$meV$), the prominent low- frequency structure at about 10-20~$meV$ (in Ho- compound it ends at about 20~$meV$ and in Dy-compound at $\sim 16\ meV$) and a flat peak at about $\sim 25\ meV$. The lowest-lying peak is presumably associated with the magnetic order at least at temperatures below AFM transition. First, in $\rm HoNi_2B_2C$ the intensity of similar peak strongly depends on the magnetic field (Fig.\ref{Fig7}).\footnote{We neglect the different magnetic structures in a field that kills superconductivity of Dy- and Ho-compounds.}  Second, this peak is very similar
to the AFM structure in Ho-compound (see Figs.9 and 10 of Ref.\cite{1}).

In Fig.\ref{Fig7}, the initial part of the PC spectrum for $\rm HoNi_2B_2C$ is replotted from Fig.7 of Ref.\cite{8}, as a function of magnetic field parallel to the $ab$ plane. The linearly rising background is subtracted from the spectra. We disregard the strong undershooting for phonon modes at $eV = 10-20\ meV$ and field 0.5 and 1.0~$T$ which is due to the quasi-thermal effect associated with the destruction of the magnetic order. Note, that for fields 2-9~$T$ the broad phonon band at $eV = 10-20\ meV$ is the same within the experimental accuracy. The Gaussian lowest-lying mode at $eV = 5\ meV$ looks like a conventional PCS-peak. Its
intensity is strongly dependent on field just in the initial range (0-2~$T$) where the magnetic order still exists and saturates at large fields (see inset of Fig.\ref{Fig7}). We emphasize that its position on the energy scale has nothing to do with superconductivity, which is completely destroyed at these fields, but probably is due to the magnetic order which changes with field.

\begin{figure}[]
\includegraphics[width=8.5cm,angle=0]{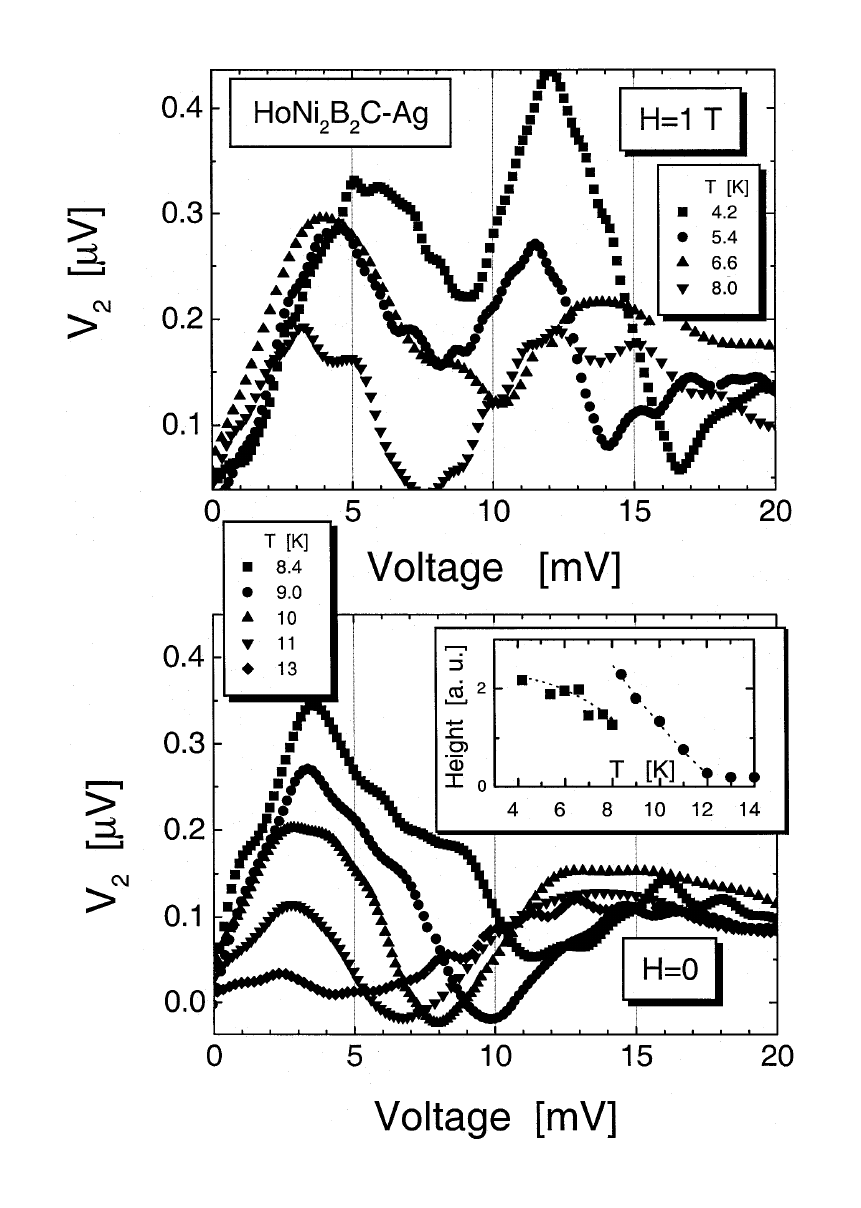}
\caption[]{PC spectra at different temperatures for $\rm HoNi_2B_2C-Ag$ contact replotted from Fig.9 of Ref.\cite{8} linear background being subtracted. For the upper and lower panels, the external magnetic field equals 1 and 0 $T$, respectively. In the inset, the height of the lowest-lying peak vs. temperature is shown, the jump around 8~$K$ is due to the change in magnetic field which influences the magnetic order. Parameters of the contact: $R_0 = 0.62\ \Omega$, $V_{1,0} = 0.95\ mV$.}
\label{Fig8}
\end{figure}

To compare the temperature dependences of $\rm HoNi_2B_2C$ and $\rm DyNi_2B_2C$, in Fig.\ref{Fig8} we show the spectra of $\rm HoNi_2B_2C-Ag$ junction, taken from Fig. 9 of Ref.\cite{8}. There are two subsets of spectra (see upper and lower panels of Fig.\ref{Fig8}). One is taken at low temperature but with a magnetic field of 1~$T$ in order to suppress superconductivity. The second is recorded at zero field but at temperatures higher than $T_c$. As the intensity of the lowest-lying peak, we take its height whose temperature dependence is plotted in the inset of the lower panel. It is seen that the height starts to increase at $T_m^* = 12\ K$ and saturates below 6~$K$. A drop at about 8~$K$ is due to the application of an external magnetic field. The onset temperature is noticeably higher than the magnetic transition temperature observed by neutrons \cite{3} ($T_m = 8.5\ K$). As it happens in $\rm DyNi_2B_2C$, there are no dramatic changes either at $T_m$ nor at $T_N = 6\ K$. Thus we see that for $\rm HoNi_2B_2C$ the lowest-lying peak behaves in the similar way to that observed for the $\rm DyNi_2B_2C$ compound.

Up to now we have not discussed the influence of CEF excitation. CEF excitations in ferromagnetically ordered $\rm NdNi_5$ were studied by PCS as a function of temperature in Ref.\cite{24}. The reported behavior is very similar to what we have observed. The appearance of CEF spectral lines at the onset temperature was explained as a decreasing population of CEF energy levels with decreasing temperature which enhances the inelastic transitions between levels \cite{25}.

For Ho-compound CEF energy levels are investigated in Refs. \cite{13,14}. No CEF magnetic scattering intensity was found above 20~$meV$ by neutron experiments \cite{14}, which coincides with main group of peaks for this compound in PC spectrum (Fig.\ref{Fig6}). Comparing with the corresponding group in $\rm DyNi_2B_2C$, the cutoff of the main group of peaks is observed at noticeably lower energy ($\sim 16\ meV$) which might be hardly explained by phonon excitations since Ho and Dy have almost the same masses.
The difference in energy positions for the lowest-lying peaks (Fig.\ref{Fig6}) also might be ascribed to the influence of CEF excitations.

Unfortunately, neither the CEF energy level scheme extracted from fitting procedure using temperature dependent magnetic susceptibility \cite{13}, nor the direct inelastic neutron scattering experiments \cite{14} were able to find in $\rm HoNi_2B_2C$ the CEF transition energy of the order of 3-4~$meV$, although such a transition was found in $\rm TmNi_2B_2C$  \cite{14}.\footnote{The AFM structure of $\rm TmNi_2B_2C$ compound is similar to Ho- and Dy-compound with ferromagnetic ordering in every single $ab$ plane. The difference is that in Tm, the magnetic moments are oriented along $c$-axis, whereas in Ho- and Dy-compound along the $ab$ plane.}

At the same time, several properties point out that the observed lowest-lying peak is not due to pure phonon excitations. Its intensity and position on the energy scale depend on the temperature and magnetic field (Figs.\ref{Fig4}, \ref{Fig5}, \ref{Fig7}, \ref{Fig8}) which is non-typical for phonon excitations. Also, the linear background in the PC spectra is not characteristic for scattering by nonequilibrium phonon excitations but looks more like electron-electron interactions \cite{8}.

Thus, we may infer that the lowest-lying peak in the EQI spectral function is influenced by the CEF magnetic correlations appearing at $T_m^*\simeq 15$ and 12~$K$ for $\rm DyNi_2B_2C$ and $\rm HoNi_2B_2C$, respectively, and not yet observed by neutron scattering \cite{3,14}.

In a recent report, Takeya and Kuznietz \cite{26} observe a small step in dc-magnetization curves for $\rm (Pr_{1-x}Dy_x)Ni_2B_2C$ compounds which for $\rm DyNi_2B_2C$ gives the temperature of magnetic transition at about $T_m^* = 16\ K$. They suggest that at this temperature the ferromagnetic ordering of Dy moments occurs by a second order phase transition in each DyC-layer separately. On lowering temperature, a 3D first order AFM transition occurs at $T_N = 10.5\ K$. At this temperature, each ferromagnetically ordered DyC layers are locked antiferromagnetically along the $c$-axis in a commensurate fashion. Their $T_m^*$ magnetic transition for $\rm DyNi_2B_2C$ \cite{26} is close to $T_m^*$ observed by us in Fig.\ref{Fig4}. On our opinion, there is rather a degree of anisotropy in the magnetic correlations. So one gets stronger exchange in the planes producing 2D like correlations and 2D-3D transition at 10.6~$K$. This would not really have a phase transition at 15~$K$ though the effects are dynamic, not static.

By analogy, we may assume that for $\rm HoNi_2B_2C$ the corresponding temperature is equal to $T_m^* = 12\ K$ for magnetic 2D correlations of the Ho-moments within each $ab$ plane separately. Since it is very hard to see the modulated magnetic structure above 6~$K$ in Ho-compounds, the same effect might exists in $\rm DyNi_2B_2C$ as well. It would have an AF component and the moment rotating on a cone about it, hence the extended Bragg intensity might be searched above 10.6~$K$ in $\rm DyNi_2B_2C$.

Let us speculate how the lowest-lying peak in the EQI function appears. We rule out the electron-magnon interaction as a source of 5~$meV$ peak at temperatures higher than $T_N$ . The neutron studies of the nonmagnetic compounds $\rm YNi_2B_2C$ and $\rm LuNi_2B_2C$ \cite{5,6,11} clearly show that there is a softening of phonon branches at the wave vector of $\simeq(0.5,0,0)$ where the Fermi surface has a nesting feature \cite{4}. This anisotropic anomalous softening leads to the formation of a new strong and narrow phonon mode at $eV = 4\ meV$ due to nonadiabatic effects \cite{27} which is seen in PC spectra \cite{7}. Its origin comes from the formation of Cooper pairs at the superconducting transition and the opening of the superconducting energy gap in these compounds.

One might speculate that in $\rm DyNi_2B_2C$ and $\rm HoNi_2B_2C$ we observe the appearance of the analogous phonon mode which is due to a gap (or quasigap) in the electron DOS caused by the short range magnetic order. Clearly, the superconducting transition which serves as a base for previous theories \cite{11,12,27,28} cannot be invoked as an explanation. The superconducting transition temperature in $\rm DyNi_2B_2C$ is far below the characteristic temperature $T_m^*$ where this mode appears. Also, the superconducting energy gaps both for $\rm DyNi_2B_2C$ and $\rm HoNi_2B_2C$ are around $2\Delta_0\simeq 2\ meV$ which is noticeably less than the central energy of this low-lying peak. On the other hand, for a magnetic transition temperature of $12\div 15\ K$ the mean field value of the magnetic energy gap in the electronic DOS fits well the observed position of the low-frequency phonon mode. Unfortunately, although the appearance of a gap (or quasi gap) in electron DOS near Fermi energy should be felt by PCS as a strong nonlinearity near zero bias, it is not observed at all.

Thus we can only guess that the reason for strong lowest-lying peaks in PC spectra of $\rm DyNi_2B_2C$ and $\rm HoNi_2B_2C$ is the interaction of conduction electrons with coupled CEF-phonon excitations (see, for example Ref.\cite{29}). The branch-crossing of low frequency phonons and CEF energy levels plus the strong interaction of conduction electrons with CEF excitations makes our peaks visible whereas the inelastic neutron magnetic scattering does not feel them.

The strong EQI in $\rm RNi_2B_2C$ materials along definite direction leads to an anisotropy of the energy gap which has not yet been observed in experiment. An inelastic scattering and pair breaking effect in anisotropic superconductor may reduce the gap-to-critical-temperature ratio to the weak coupling value \cite{30}. This may explain why the electron-phonon structure of superconducting $\rm RNi_2B_2C$ has not yet been observed in the tunneling experiments \cite{31}. Further work is needed to elucidate this issue.

\section{Conclusion}
We have found a strong and narrow low-frequency peak ($eV\sim 5\ meV$) in the EQI function in $\rm DyNi_2B_2C$ below $T_m^* = 15\ K$. The low-frequency peak shows no change (in position and intensity) at the magnetic transition temperatures ($T_N = 10.5$ and 6~$K$ for $\rm DyNi_2B_2C$ and $\rm HoNi_2B_2C$, respectively, and $T_m = 8.5\ K$ for $\rm HoNi_2B_2C$) suggesting that its origin is due not to the spin-density wave excitations (magnons), but most probably to phonons coupled with CEF magnetic excitons.

This low frequency peak plays a crucial role in determining the magnitude of the EQI parameter, $\lambda$, since it is located at a very low energy. A more detailed account of this peak in the superconductivity mechanism would require a closer theoretical inspection of the effects discussed here.
\section{Acknowledgements}
I.K.Y. is grateful for financial support from the International Science Foundation (Soros Foundation).
The authors acknowledge illuminating discussions on CEF PCS with M. Reiffers.


\begin{thebibliography}{}


\bibitem{1}I.K. Yanson, N.L. Bobrov, C.V. Tomy, D.McK. Paul. Point-contact spectroscopy of superconducting energy gap in $\rm DyNi_2B_2C$, \href{http://dx.doi.org/10.1016/S0921-4534(00)00227-6}{Physica C}: \textbf{334}, 2000, 33-43; \href{https://arxiv.org/pdf/1702.05785.pdf}{arXiv:1702.05785}.
\bibitem{2}T. Terashima, C. Haworth, H. Takeya, S. Uji, H. Aoki, K. Kadowaki, \href{https://doi.org/10.1103/PhysRevB.56.5120}{Phys. Rev. B} \textbf{56} (1997) 5120.
\bibitem{3}	J.W. Lynn, S. Skanthakumar, Q. Huang, S.K. Sinha, Z. Hossain, L.C. Gupta, R. Nagarajan, C. Godart, \href{https://doi.org/10.1103/PhysRevB.55.6584}{Phys. Rev. B} \textbf{55} (1997) 6584.
\bibitem {4}J.Y. Rhee, X. Wang, B.N. Harmon, \href{https://doi.org/10.1103/PhysRevB.51.15585}{Phys. Rev. B} \textbf{51} (1995) 15585.
\bibitem{5}	H. Kawano, H. Yoshizawa, H. Takeya, K. Kadowaki, \href{https://doi.org/10.1103/PhysRevLett.77.4628}{Phys. Rev. Lett.} \textbf{77} (1996) 4628.
\bibitem{6}	C. Stassis, M. Bullock, J. Zaretsky, P. Canfield, A. Goldman, G. Shirane, S.M. Shapiro, \href{https://doi.org/10.1103/PhysRevB.55.R8678}{Phys. Rev. B} \textbf{55} (1997) 8678.	
\bibitem{7}I.K. Yanson, V.V. Fisun, A.G.M. Jansen, P. Wyder, P.C. Canfield, B.K. Cho, C.V. Tomy, D.McK. Paul, \href{https://doi.org/10.1103/PhysRevLett.78.935}{Phys. Rev. Lett.} \textbf{78} (1997) 935.
\bibitem{8}	I.K. Yanson, V.V. Fisun, A.G.M. Jansen, P. Wyder, P.C. Canfield, B.K. Cho, C.V. Tomy, D.McK. Paul, \href{http://fntr.ilt.kharkov.ua/fnt/pdf/23/23-9/f23-0951e.pdf}{Fiz. Nizk. Temp.} \textbf{23} (1997) 951, [\href{http://dx.doi.org/10.1063/1.593378}{Low Temp. Phys.} \textbf{23} (1997) 712].
\bibitem{9}	I.K. Yanson, in: M. Ausloos, S. Kruchinin (Eds.), \href{https://books.google.com.ua/books?id=BCIBI7HyAsgC&pg=PA271&lpg=PA271&dq=I.K.+Yanson,+in:+M.+Ausloos,+S.+Kruchinin+(Eds.),
+Symmetry+and+Pairing+in+Superconductors,+Kluwer+Academic+Publishing,+1999,+pp.+271-285&source=bl&ots=PoCiV0C0tU&sig=z9Fqnd9fjBEft3AnaSusxgxBt4w&hl
=ru&sa=X&ved=0ahUKEwir2oGvnpvSAhXIkSwKHWfjDCsQ6AEIHDAB#v=onepage&q=I.K.}{Symmetry and Pairing in Superconductors}, Kluwer Academic Publishing, 1999, pp. 271-285.

\bibitem{10}F. Gompf, W. Reichardt, H. Schober, B. Renker, M. Buchgeister, \href{https://doi.org/10.1103/PhysRevB.55.9058}{Phys. Rev. B} \textbf{55} (1997) 9058.
\bibitem{11}M. Bullock, J. Zarestky, C. Stassis, A. Goldman, P. Canfield, Zentaro Honda, Gen Shirane, and S. M. Shapiro, \href{https://doi.org/10.1103/PhysRevB.57.7916}{Phys. Rev. B} \textbf{57} (1998) 7916.
\bibitem{12}P.B. Allen, V.N. Kostur, N. Takesue, G. Shirane, \href{https://doi.org/10.1103/PhysRevB.56.5552}{Phys. Rev. B} \textbf{56} (1997) 5552.	
\bibitem{13}B.K. Cho, B.N. Harmon, D.C. Johnston, P.C. Canfield, \href{https://doi.org/10.1103/PhysRevB.53.2217}{Phys. Rev. B} \textbf{53} (1996) 2217.	
\bibitem{14} U. Gasser et al., \href{https://doi.org/10.1007/BF02583718}{Czech. J. Phys.} \textbf{46} (Suppl. S2) (1996). in: Proceedings of the 21st International Conference on Low Temperature Physics, Prague, August 8-14,1996, p. 821.	
\bibitem{15}A.V. Khotkevich, I.K. Yanson, \href{https://books.google.com.ua/books?hl=ru&lr=&id=yLT4BwAAQBAJ&oi=fnd&pg=PR7&ots=nx4U1iBBTd&sig=vObbXR0S9m8NfCFXT2rtQTvhiq0&redir_esc=y#v=onepage&q&f=false}{Atlas of Point Contact Spectra of Electron-Phonon Interactions in Metals}, Kluwer Academic, New York, 1995.	
\bibitem{16}A.K. Bhatnagar, K.D.D. Rathnayaka, D.G. Naugle, P.C. Canfield, \href{https://doi.org/10.1103/PhysRevB.56.437}{Phys. Rev. B} \textbf{56} (1997) 437.
\bibitem{17} G.E. Blonder, M. Tinkham, T.M. Klapwijk, \href{https://doi.org/10.1103/PhysRevB.25.4515}{Phys. Rev. B} \textbf{25} (1982) 4515.
\bibitem{18}Z.Q. Peng, K. Krug, K. Winzer, \href{https://doi.org/10.1103/PhysRevB.57.R8123}{Phys. Rev. B} \textbf{57} (1998) R8123.	
\bibitem{19}L.F. Mattheiss, T. Siegrist, J. Cava, \href{http://dx.doi.org/10.1016/0038-1098(94)90551-7}{Solid State Commun.} 91 (1994) 587.
\bibitem{20}L.F. Rybaltchenko et al., \href{http://iopscience.iop.org/article/10.1209/epl/i1996-00367-8/meta}{Europhys. Lett.} \textbf{33} (1996) 483.	
\bibitem{21}I.K. Yanson, V.V. Fisun, N.L. Bobrov, L.F. Rybal'chenko, \href{http://www.jetpletters.ac.ru/ps/1244/article_18813.pdf}{JETP Lett.} \textbf{45} (1987) 543; \href{https://arxiv.org/pdf/1602.04356v1.pdf}{arXiv:1602.04356}.	
\bibitem{22}V.A. Khlus, \href{http://fntr.ilt.kharkov.ua/fnt/pdf/9/9-9/f09-0985r.pdf}{Fiz. Nizk. Temp.} \textbf{9} (1983) 985, [Sov. J. Low Temp. Phys. \textbf{9} (1983) 510].	
\bibitem{23}I.K. Yanson, G.V. Kamarchuk, A.V. Khotkevich, \href{http://fntr.ilt.kharkov.ua/fnt/pdf/10/10-4/f10-0423r.pdf}{Fiz. Nizk. Temp.} \textbf{10} (1984) 423, [Sov. J. Low Temp. Phys. \textbf{10} (1984) 220].	
\bibitem{24}M. Reiffers, T. Salonova, D. Gignoux, D. Schmitt, \href{https://doi.org/10.1209/epl/i1999-00197-2}{Europhys. Lett.} \textbf{45} (1999) 520, See, also the references cited therein.	
\bibitem{25}M. Reiffers, private communication.
\bibitem{26}H. Takeya, M. Kuznietz, \href{http://dx.doi.org/10.1016/S0921-4526(98)01006-0}{Physica B} \textbf{259-261} (1999) 596.
\bibitem{27}A.E. Karakozov, E.G. Maksimov, \href{http://www.jetp.ac.ru/cgi-bin/dn/r_115_1799.pdf}{Zh. Eksp. Teor. Fiz.} \textbf{115} (1999) 1799.	
\bibitem{28}H.-Y. Kee, C.M. Varma, \href{https://doi.org/10.1103/PhysRevLett.79.4250}{Phys. Rev. Lett.} \textbf{79} (1997) 4250.	
\bibitem{29}P. Fulde, in: K.A. Gschneidner Jr., L. Eyring (Eds.), \href{http://www.fulviofrisone.com/attachments/article/426/Handbook\%20on\%20the\%20Physics\%20and\%20Chemistry\%20of\%20Rare\%20Earths\%20Volume\%2022.pdf}{Handbook on the Physics and Chemistry} of Rare Earths 2 North-Holland, Amsterdam, 1978, Chap. 17.
\bibitem{30}A.J. Millis, S. Sachdev, C.M. Varma, \href{https://doi.org/10.1103/PhysRevB.37.4975}{Phys. Rev.B}  \textbf{37} (1988) 4975.	
\bibitem{31}T. Ekino et al., \href{https://doi.org/10.1103/PhysRevB.53.5640}{Phys. Rev. B} \textbf{53} (1996) 5640.

\end{thebibliography}
\end{document}